\documentclass[showpacs,twocolumn]{revtex4}

\usepackage{ragged2e}
\usepackage{amsmath}
\usepackage{graphicx}
\usepackage{rotating}

\usepackage[usenames,dvipsnames]{color}

\definecolor{darkblue}{RGB}{0,0,196}

\begin{document}

\title{\Large \bf Initial-state temperature of light meson emission source
from squared momentum transfer spectra in high-energy
collisions\vspace{0.5cm}}

\author{Qi Wang$^{1,}$\footnote{qiwang-sxu@qq.com;
18303476022@163.com}, Fu-Hu Liu$^{1,}$\footnote{Correspondence:
fuhuliu@163.com; fuhuliu@sxu.edu.cn}, Khusniddin K.
Olimov$^{2,}$\footnote{Correspondence: khkolimov@gmail.com;
kh.olimov@uzsci.net}}

\affiliation{$^1$Institute of Theoretical Physics, State Key
Laboratory of Quantum Optics and Quantum Optics Devices \&
Collaborative Innovation Center of Extreme Optics, Shanxi
University, Taiyuan 030006, China
\\
$^2$Laboratory of High Energy Physics, Physical-Technical
Institute of Uzbekistan Academy of Sciences, Chingiz Aytmatov str.
$2^b$, 100084 Tashkent, Uzbekistan}

\begin{abstract}

\vspace{0.5cm}

\noindent {\bf Abstract:} The squared momentum transfer spectra of
light mesons, $\pi^0$, $\pi^+$, $\eta$, and $\rho^0$, produced in
high-energy virtual photon-proton ($\gamma^{*} p$) $\rightarrow
{\rm meson + nucleon}$ process in electron-proton ($ep$)
collisions measured by the CLAS Collaboration are analyzed by the
Monte Carlo calculations, where the transfer undergoes from the
incident $\gamma^*$ to emitted meson or equivalently from the
target proton to emitted nucleon. In the calculations, the Erlang
distribution from a multi-source thermal model is used to describe
the transverse momentum spectra of emitted particles. Our results
show that the average transverse momentum ($\langle p_T\rangle$)
and the initial-state temperature ($T_i$) increase from lower
squared photon virtuality ($Q^2$) and Bjorken variable ($x_B$) to
higher one. This renders that the excitation degree of emission
source, which is described by $\langle p_T\rangle$ and $T_i$,
increases with increasing of $Q^2$ and $x_B$.

\end{abstract}
\pacs{12.40.Ee, 14.40.-n, 24.10.Pa, 25.75.Ag \vspace{0.5cm}}

\maketitle

\section{Introduction}

In the evolution process of high-energy nucleus-nucleus
(heavy-ion) collisions, the reaction system undergoes several main
stages which are separately the incoming of nuclei, beginning of
collisions, strongly-coupled quark-gluon plasma (sQGP) phase or
hot-dense matter phase, mixed phase, and hadron gas. In the stage
of the incoming of nuclei, two nuclei move toward each other in
vacuum tunnel at nearly the speed of light and change the shape to
pancake with the Lorentz contraction. The sQGP phase is extremely
hot-dense matter and the system can be regarded as a fireball.
Considering the effect of pressure gradient, the system begins to
inflate and cool down. Then, the hadron matter appears until the
system is hadronic. To understand the mechanism of nuclear
reaction and the property of system evolution, it is necessary to
investigate the characteristics of each stage of collision
process. The excitation and equilibrium degrees of the system are
among very important
characteristics~\cite{1a,1b,1c,1d,1e,1f,1g,1h,1i,1j}.

To describe the excitation degree of the system, various
temperatures of the system and the average transverse momentum
($\langle p_T\rangle$) of particles are
used~\cite{1,2,3,4,5,6,7,8}. The various temperatures include, but
are not limited to, i) the initial-state temperature ($T_i$) which
reflects the temperature in the beginning of collisions of two
nuclei, ii) the chemical freeze-out temperature ($T_{ch}$) which
reflects the temperature at chemical freeze-out when inelastic
collisions disappear, iii) the kinetic freeze-out or final-state
temperature ($T_{kin}$ or $T_0$) which reflects the temperature at
kinetic freeze-out when elastic collisions disappear, and iv) the
effective temperature ($T_{eff}$) which is not a ``real"
temperature, in which the influence of flow effect is not excluded
compared with $T_{kin}$ or $T_0$. Different kinds of temperatures
can be ``measured" by different ``thermometers" (methods).

As the earliest temperature in collisions, $T_i$ is used to
explore the secret of high-energy collisions~\cite{1,2,3,4,5}. As
we know, $T_i$ is the temperature of emission source or
interacting system when the system undergoes the initial-stage of
collisions~\cite{9}. It is interesting for us to describe the
excitation degree of the system by using $T_i$. Generally, from
the transverse momentum ($p_T$) spectra or fitting the $p_T$
spectra with different distributions or functions, we may obtain
$T_i$. The Erlang distribution~\cite{10,11,12}, Tsallis
distribution~\cite{12a,12b}, Hagedorn function~\cite{12c} are
usually used, but in this paper, we only choose the Erlang
distribution due to its origin of multiple sources in the
multi-source thermal model~\cite{10,11,12}. In the special case,
such as absent $p_T$ spectra, the squared momentum transfer
spectra are alternatively used. Obviously, $T_i$ can not be
obtained from the squared momentum transfer spectra directly
unless the $p_T$ spectra are transformed to them. From the fit to
$p_T$ spectra, $\langle p_T\rangle$ can be naturally abstracted.

In the transformation of $p_T$ spectra to squared momentum
transfer spectra~\cite{13}, the Monte Carlo method is used. First
of all, concrete $p_T$, satisfying the Erlang
distribution~\cite{10,11,12}, are produced. Then, the squared
momentum transfers are calculated according to the relation
between squared momentum transfers and $p_T$ by using the Monte
Carlo method. At last, the distribution of squared momentum
transfer spectra are obtained and used to fit the experimental
data for extracting $\langle p_T\rangle$ and $T_i$.

To describe the equilibrium degree of the system, one can use the
Tsallis distribution~\cite{12a,12b} or Hagedorn
function~\cite{12c} to fit $p_T$ spectra directly. In the fitting
process, the entropy index $q$ can be extracted. The closer to 1
the entropy index $q$ is, the higher the degree of equilibrium of
the source or system is. The relation between the two
distributions is that the former one covers the later one in which
the mass is neglected. Because the universality, similarity, or
common characteristics exist in high-energy
collisions~\cite{13a,13b,13c,13d,13e,13f,13g,13h,13i,13j}, some
distributions used in large collision system can be also used in
small collision system. Although the equilibrium degree is also
important, it is not discussed in this work due to other topics
being concerned. We think that the equilibrium degree is enough to
use the concept of temperature.

Meson consists of a quark and anti-quark ($q\bar q$) and belongs
to hadron. It takes part in the strong interaction and play an
important role. Light meson refers to a kind of meson with low
mass. The transverse momentum of light meson changes more
sensitively than that of the heavy one. Therefore, the study of
transverse momentum spectra of light mesons is very important to
explore the reaction mechanism and evolution process of
high-energy collisions.

Compared with large systems of high-energy nucleus-nucleus
collisions, small systems such as high-energy electron-proton,
proton-proton, proton-nucleus collisions also produce abundant
results. In particular, in electron-proton collisions, the
scattered electron exchanges virtual photon ($\gamma$$^{*}$) with
the target proton. Then, one may study high-energy $\gamma$$^{*}$
induced proton collisions, that is $\gamma$$^{*} p$ collisions,
experimentally, theoretically as well as phenomenologically.

In this paper, the squared momentum transfer spectra of light
mesons, $\pi^0$, $\pi^+$, $\eta$, and $\rho^0$, produced in
high-energy $\gamma$$^{*} p$ collisions measured by the CLAS
Collaboration~\cite{14,15,16,17} are fitted by the results
originating from the Erlang $p_T$ distribution with the Monte
Carlo method. The CLAS experimental data are measured at different
squared photon virtuality $Q^2$ and Bjorken variable $x_B$, where
$Q^2$ and $x_B$ will be discussed later in the subsection 2.3.

\section{Formalism and method}

i) {\it The Erlang distribution}

The Erlang distribution is a direct result of the multi-source
thermal model~\cite{10,11,12}. One or two-component Erlang
distribution can describe the narrow or wide $p_T$ spectra of
particles, where the narrow (wide) $p_T$ spectra refers to range
less than a few GeV/$c$ (more than 10 GeV/$c$)~\cite{12}. The
multi-source thermal model assumes that multiple sources are
formed in high-energy collisions. These sources can be nucleons or
partons if we study the formation of nucleon clusters (nuclear
fragments) or particles.

In this work, we assume that a few ($n_s$) partons (partons-like)
contribute to $p_T$ of a given particle~\cite{12}. The
contribution of the $j$-th parton is assumed to be an exponential
function with variable $p_{tj}$ which depends on $j$, and average
value $\langle p_t\rangle$ which is independent of $j$. We have
the normalized exponential function
\begin{align}
f(p_{tj})=\frac{1}{\langle p_t \rangle} \exp\bigg(-\frac{p_{tj}}{\langle
p_t \rangle}\bigg).
\end{align}
Here, $\langle p_t\rangle$ represents the average contribution of
participant partons to $\langle p_T\rangle$ of the considered
particles.

The contribution sum ($p_{t1}+p_{t2}+...+p_{tn_s}$) of $n_s$
partons is $p_T$ of a given particle. The result convoluting the
contributions of $n_s$ partons is the Erlang distribution. We have
the Erlang $p_T$ distribution to be
\begin{align}
f(p_T)=\frac{1}{N}\frac{dN}{dp_T}=\frac{p_T^{n_s-1}}{(n_s-1)!{\langle
p_t \rangle}^{n_s}} \exp\bigg(-\frac{p_T}{{\langle p_t
\rangle}}\bigg).
\end{align}
Here $N$ is the number of particles, and the form of
$(1/N)dN/dp_T$ results in the normalization of $f(p_T)$ to 1. In
fact, the normalization of the Erlang distribution is naturally 1.

We would like to emphasize here the difference between ``$n_s$",
the number of partons and ``$N$", the number of particles. In
$\gamma^* p$ collisions, if three quarks in the proton contributed
to $p_T$, we have $n_s=3$. If another $q\bar q$ pair also
contributed to $p_T$, we have $n_s=3+2=5$. Even in nucleus-nucleus
collisions, the value of $n_s$ is not large due to it being
determined by the number of contributor partons in a
nucleon-nucleon pair, but not collision system itself. This makes
sense, in the Fock's first two terms of the development of the
wave function of the proton, as composed by 3 quarks and then 3
quarks plus a $q\bar q$ pair~\cite{17a,17b}. As for $N$, its value
may be small in small collision system or peripheral
nucleus-nucleus collisions. The value of $N$ may be very large in
central nucleus-nucleus collisions at high energy.
\\

ii) {\it Average transverse momentum and initial-state
temperature}

As we know, both the average transverse momentum $\langle
p_T\rangle$ and initial-state temperature $T_i$~\cite{1,2,3,4,5}
describe the excitation degree of the system. In particular, in
the Erlang distribution, $\langle p_T\rangle$ can be easily
obtained by
\begin{align}
\langle p_T\rangle =\int_0^{\infty} p_T f(p_T) dp_T =n_s\langle
p_t\rangle,
\end{align}
where $f(p_T)$ is normalized to 1. Similarly, $\langle p_t\rangle$
reflects the excitation degree of participant partons.

According to refs.~\cite{18,19,20}, with a color string
percolation method~\cite{21}, $T_i$ can be regarded as
\begin{align}
T_i =\sqrt{\frac{\langle p_T^2 \rangle}{2F(\xi)}},
\end{align}
where
\begin{align}
\langle p_T^2\rangle =\int_0^{\infty} p^2_T f(p_T) dp_T
\end{align}
due to $f(p_T)$ is normalized to 1 and $\sqrt{\langle
p_T^2\rangle}$ is the root-mean-square of $p_T$. In Eq. (4),
$F(\xi)$ is the color suppression factor~\cite{21}.

In the process of using color string method to obtain $T_i$ in
this work, only one string is used, i.e. $F(\xi)=1$, in the
formation of particle. Although there are probability to have any
other strings, they do not affect noticeably $T_i$. If we consider
other strings, according to ref.~\cite{21}, one has the minimum
$F(\xi)\approx0.6$. This will cause the maximum increase of 29.1\%
in $T_i$. Considering the fraction of one string is very large,
that of two strings is relative small, and that of multiple
strings is very small, the increase in $T_i$ will be much smaller
than 29.1\%.
\\

iii) {\it The squared momentum transfer}

In the center-of-mass reference frame, in two-body process
$2+1\rightarrow 4+3$ or two-body-like process of high-energy
collisions, there are three Mandelstam variables defined based on
the four-momenta of these particles. They have the forms to be
\begin{align}
s=-({P_1}+{P_2})^{2}=-({P_3}+{P_4})^{2},
\end{align}
\begin{align}
t=-({P_1}-{P_3})^{2}=-(-{P_2}+{P_4})^{2},
\end{align}
\begin{align}
u=-({P_1}-{P_4})^{2}=-(-{P_2}+{P_3})^{2},
\end{align}
where $P_{1}$, $P_{2}$, $P_{3}$, and $P_{4}$ are four-momenta of
particles 1 (target proton), 2 (incident $\gamma^*$), 3 (emitted
nucleon), and 4 (emitted meson), respectively. Here, we assume
that particle 1 is incident along the $Oz$ direction and particle
2 is incident along the opposite direction. After collisions,
particle 3 is emitted with angle $\theta$ relative to the $Oz$
direction and particle 4 is emitted along the opposite direction.

The three Mandelstam variables have different physical meaning.
For instance, $\sqrt{s}$ refers to the center-of-mass energy, and
$-u$ is defined as the squared momentum transfer between particles
1 and 4. Here, selected variable $-t$ (the squared momentum
transfer between particles 1 and 3) is calculated to fit the
experimental data. For convenience, we have
\begin{align}
|t|=& |({E_1}-{E_3})^{2}-({\vec{p}_{1}}-{\vec{p}_{3}})^{2}| \nonumber\\
=& \bigg|m_1^2+m_3^2-2{E_1}\sqrt{\bigg({\frac{p_{3T}}{\sin\theta}}\bigg)^{2}+m_3^2} \nonumber\\
& +2\sqrt{E_1^2-m_1^2}\frac{p_{3T}}{\tan\theta}\bigg|,
\end{align}
where $E_1$ and $E_3$, $\vec{p}_1$ and $\vec{p}_3$, as well as
$m_1$ and $m_3$ are the energy, momentum, and rest mass of
particles 1 and 3, respectively. In addition, $p_{3T}$ referred to
be perpendicular to the $Oz$ direction component of the transverse
momentum of particle 3, which obeys the Erlang distribution, that
is Eq. (2) in which $p_T=p_{3T}$.

In this paper, the squared momentum transfer spectra of light
meson at different squared photon virtuality $Q^2$ and Bjorken
variable $x_B$ are fitted by calculated results with the Monte
Carlo method. Here, $Q^2$ is a reflection of hard scale of
reaction~\cite{22,23,24,25,26,27,28,29}. The harder the reaction
is, the higher the excitation degree is. In fact, $Q^2$ is the
absolute value of the squared mass of $\gamma^*$ (particle 2) that
is exchanged between the scattered electron and the target proton
(particle 1), and it effectively represents the transverse size of
the probe~\cite{15}. In addition, $-Q^2$ is also the squared
momentum transfer to the target proton (particle 1) by the
scattered electron~\cite{14}.

As for the Bjorken variable $x_B$, it represents contrarily the
momentum of particle 1. The lower the $x_B$ is, the higher the
momentum of particle 1 is. Generally, $x_B=Q^2/(2P_2\cdot\sqrt{-
Q^2})\propto Q$~\cite{14}. In a symmetric frame, importing $\xi'$
as skewness, it is half of the longitudinal momentum fraction
transferred to the struck parton. The skewness $\xi'$ can be used
to express $x_B$ approximately. That is $x_B \approx
2\xi'/(1+\xi')$~\cite{14}.
\\

iv) {\it The process of Monte Carlo calculations}

In the calculations of squared momentum transfer, the analytical
expression of $p_T$ distribution is difficult to be transformed to
that of squared momentum transfer distribution directly by using
Eq. (9). Alternatively, we may use the Monte Carlo method to
transform $p_T$ to squared momentum transfer. Let $R_{1,2}$ and
$r_{1,2,3,...,n_s}$ be random numbers distributed evenly in [0,1].
Then, many concrete transverse momentum $p_{3T}$ satisfied with
Eq. (2) and $\theta$ are produced. Other quantities such as $E_1$,
$m_1$, and $m_3$ in the equation are fixed, though $E_1$ is
treated as a parameter in the present work.

Generally, we may solve the equation
\begin{align}
\int_0^{p_T}f\left(p'_T\right)dp'_T<R_1<\int_0^{p_T+\delta
p_T}f\left(p'_T\right)dp'_T,
\end{align}
where $\delta p_T$ is a small shift relative to $p_T$.
Conveniently, there is a simpler expression due to Eqs. (1) and
(2). In fact, solving the equation
\begin{align}
\int_0^{p_{tj}}f\left(p'_{tj}\right)dp'_{tj}=r_j\qquad(j=1,2,3,...,n_s),
\end{align}
we have
\begin{align}
p_{tj}=-\langle p_t\rangle \ln r_j\qquad(j=1,2,3,...,n_s).
\end{align}
The simpler expression is
\begin{align}
p_{T}=\sum_{j=1}^{n_s} p_{tj} =-\langle p_t\rangle
\sum_{j=1}^{n_s} \ln r_j=-\langle p_t\rangle
\ln\bigg(\prod_{j=1}^{n_s} r_j\bigg).
\end{align}

The distribution of $\theta$ satisfies with the half-sine function
\begin{align}
f_\theta\left(\theta\right)= \frac{1}{2}\sin \theta
\end{align}
which is obtained under the assumption of isotropic emission in
the source's rest frame. Solving the equation
\begin{align}
\int_0^{\theta}f_\theta\left(\theta'\right)d\theta'=R_2,
\end{align}
we have
\begin{align}
\theta=2\arcsin \left(\sqrt{R_2}\right)
\end{align}
which is needed in the calculations.

We have check the consistency and correctness of the above
expressions in the Monte Carlo method in terms of illustration
which is not presented here. After obtaining concrete values of
$p_{3T}$ and $\theta$, and using $E_1$, $m_1$, and $m_3$, the
value of $|t|$ can be obtained from Eq. (9). Through repeating the
calculations many times, the distribution of $|t|$ is obtained
statistically. Based on the method of least squares, the parameter
$\langle p_t\rangle$ and $n_s$ are extracted naturally. Meanwhile,
$T_i$ can be obtained from Eq. (4) and $\langle p_T\rangle$
($\langle p_T^2\rangle$) can be obtained from Eq. (3) [(5)] or
from the statistics. The errors of parameters are obtained by the
general method of statistical analysis.

It should be noted that the above Monte Carlo calculation is only
performed in the transformation from transverse momentum to $|t|$,
in which the physics process such as the radiative corrections for
reactions induced by electrons has been taken into account
naturally. In fact, the effects of the mentioned process and all
other processes are included in the Erlang distribution which is a
result of multi-factor interactions. In other words, the Monte
Carlo calculation used here is not a simulation for the system
evolution from initial to final stages, but the numerical
transformation in the final stage.

\section{Results and discussion}

\subsection{Comparison with data}

Figure 1 shows the differential cross-section, $d\sigma/d|t|$, in
squared momentum transfer $|t|$ of $\gamma^*p \rightarrow \pi^0p$
process produced in 5.75 GeV electron beam induced collisions in a
2.5 cm long liquid-hydrogen target ($ep$ collisions at beam energy
of 5.75 GeV) in different ranges of squared photon virtuality,
$1.0<Q^2<1.5$, $1.5<Q^2<2.0$, $2.0<Q^2<2.5$, $2.5<Q^2<3.0$,
$3.0<Q^2<3.5$, $3.5<Q^2<4.0$, and $4.0<Q^2<4.6$ GeV$^2$, from
bottom to up sub-panels, as well as in different ranges of Bjorken
variable, $0.10<x_B<0.15$, $0.15<x_B<0.20$, $0.20<x_B<0.25$,
$0.25<x_B<0.30$, $0.30<x_B<0.38$, $0.38<x_B<0.48$, and 0.48
$<x_B<$ 0.58, from left to right sub-panels. The sample at the
top-left sub-panel shows repeatedly the result in the range of
squared photon virtuality, $1.5<Q^2<2.0$ GeV$^2$, and the range of
Bjorken variable, $0.20<x_B<0.25$, as an example. The symbols
represent the experimental data measured by the CLAS
Collaboration~\cite{14} and the curves are the statistical results
of squared momentum transfer $|t|$.

\begin{figure*}[htbp]
\begin{center}
\rotatebox[origin=c]{-90}{\includegraphics[width=22cm]{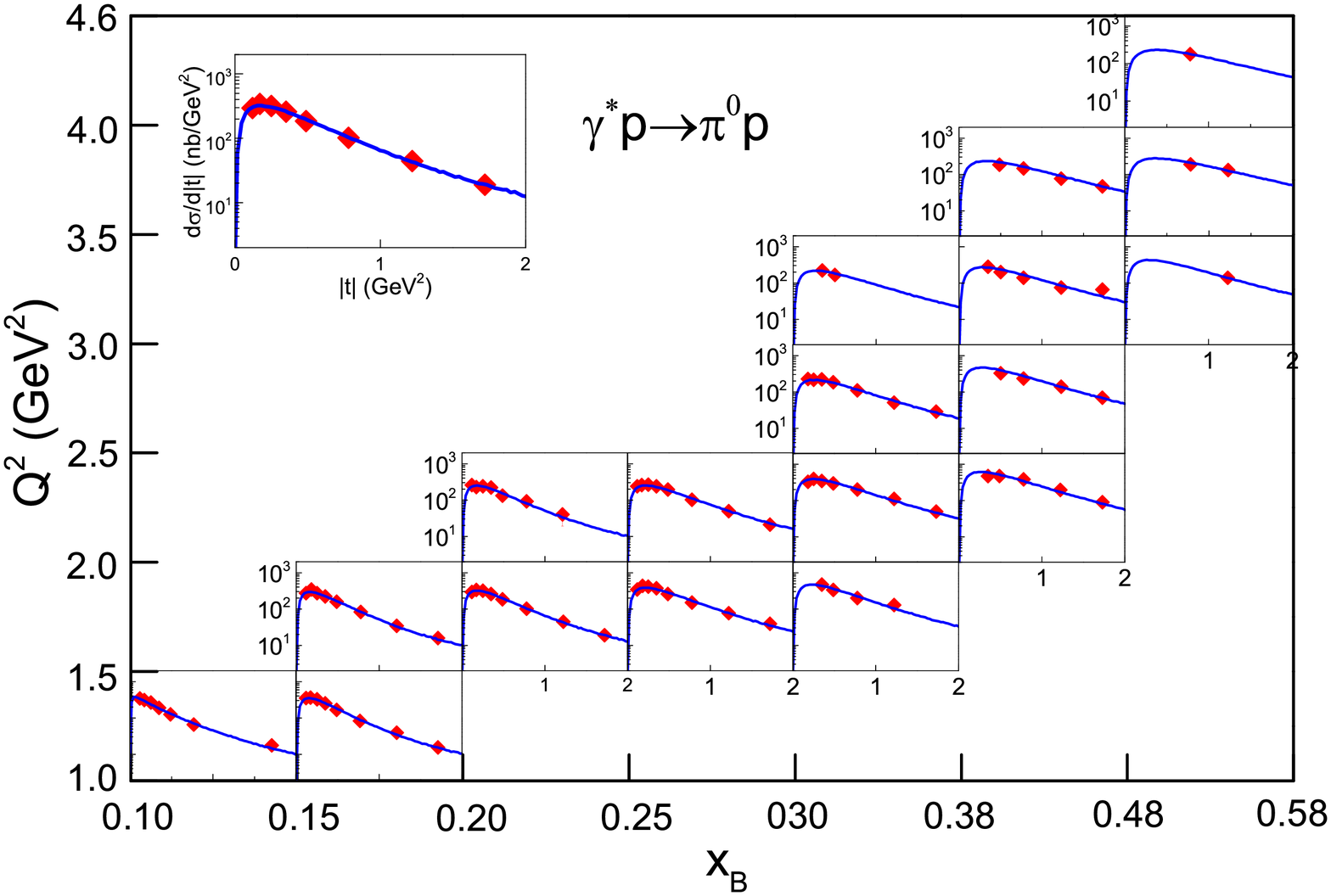}}
\end{center} \vspace{5mm}
\justifying\noindent {\small Fig. 1. The differential
cross-section $d\sigma/d|t|$ in $|t|$ of $\gamma^*p \rightarrow
\pi^0p$ process produced in $ep$ collisions at beam energy of 5.75
GeV in different ranges of $Q^2$ and $x_B$ shown in the panels.
The sample at the top-left sub-panel shows repeatedly the result
in $1.5<Q^2<2.0$ GeV$^2$ and $0.20<x_B<0.25$ as an example. The
symbols represent the experimental data measured by the CLAS
Collaboration~\cite{14} and the curves are the statistical results
of $|t|$ [Eq. (9)] in which $p_{3T}$ satisfies the Erlang
distribution [Eq. (2)] and can be obtained with the Monte Carlo
method [Eq. (13)].}
\end{figure*}

\begin{figure*}[htbp]
\begin{center}
\rotatebox[origin=c]{-90}{\includegraphics[width=20cm]{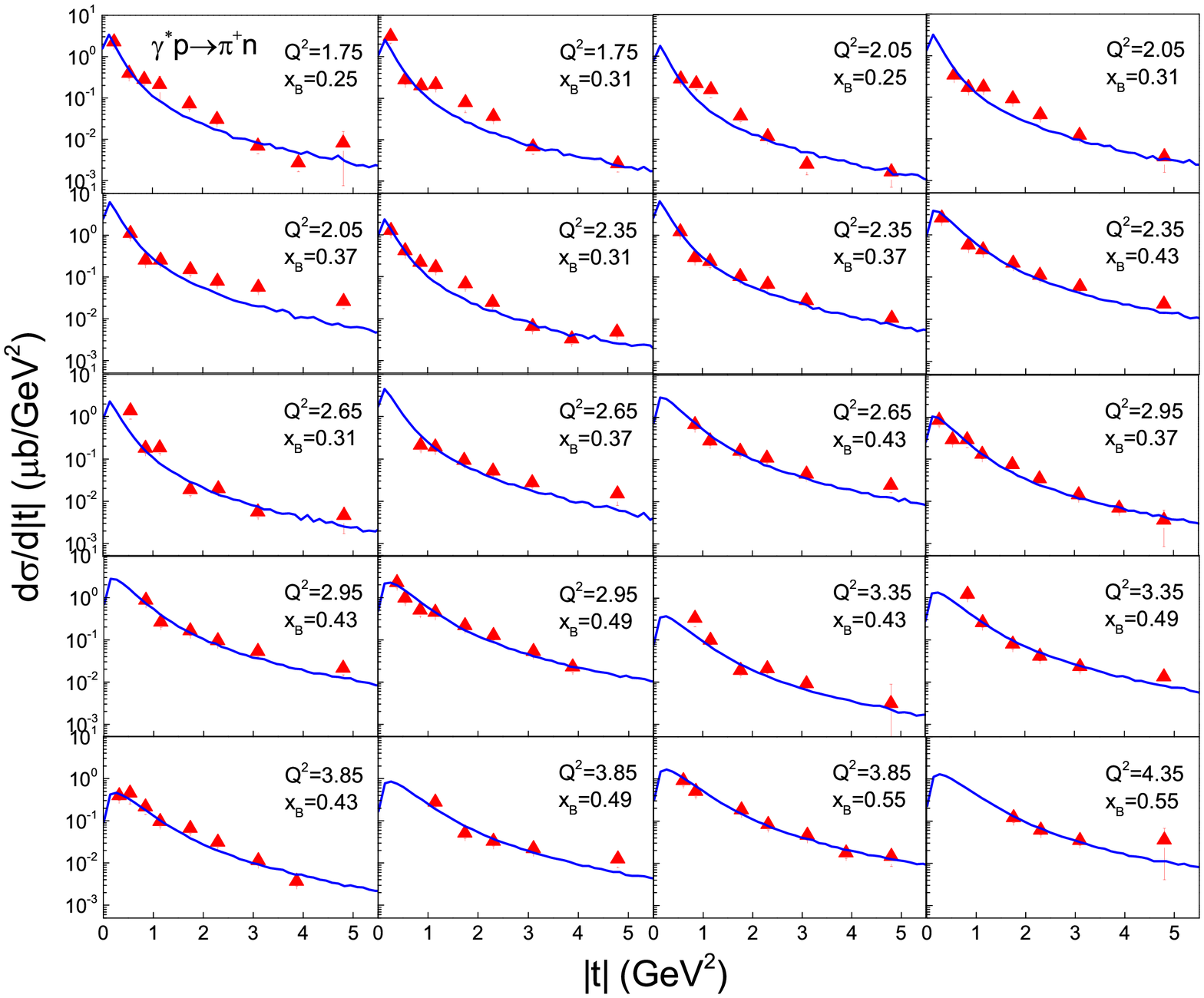}}
\end{center} \vspace{5mm}
\justifying\noindent {\small Fig. 2. The differential
cross-section $d\sigma/d|t|$ in $|t|$ of $\gamma^*p \rightarrow
\pi^+n$ process produced in $ep$ collisions at beam energy of 6
GeV at different $Q^2$ and $x_B$ shown in the panels. The symbols
represent the experimental data measured by the CLAS
Collaboration~\cite{15} and the curves are the statistical results
obtained as those in Figure 1.}
\end{figure*}

\begin{table*}[!htb]
\vspace{.0cm} \justifying\noindent {\small Table 1. Values of
$E_1$, $\langle p_t\rangle$, $n_s$, $\sigma_0$, $T_i$, and
$\chi^2$/ndof corresponding to the curves in Figures 1 and 2,
where $n_s$ is constrained to be integer with uncertainty of 0
which is not listed in the table. The number of parameters is
always 4 which includes $E_1$, $\langle p_t\rangle$, $n_s$, and
$\sigma_0$. In the case of ndof being less than or equal to the
number of parameters, we obtain the curve from a ``prediction" or
extrapolation based on other reasonable fits, and show the
corresponding nop in a bracket to replace ndof. The value of
$\chi^2$ is rounded to an integer, or one significant digit if the
integer is 0.} \vspace{-4mm}
\begin{center}
\setlength\tabcolsep{6pt}%
{\small
\begin{tabular} {cccccccccc}\\ \hline\hline  Collisions & $Q^2$ (GeV) & $x_B$ & $E_1$ (GeV) & $\langle p_t\rangle$ (GeV/$c$) & $n_s$ & $\sigma_0$ ($\mu$b) & $T_i$ (GeV) & $\chi^2$/ndof(nop)\\
\hline
$\gamma^*p \rightarrow \pi^0 p$ & $(1.0,1.5)$ & $(0.10,0.15)$ & $0.945^{+0.015}_{-0.007} $ & $0.176\pm0.002$ & $3$ & $0.195\pm0.005$ & $0.409\pm0.005$ & $7/11$\\
                                & $(1.0,1.5)$ & $(0.15,0.20)$ & $0.945\pm0.004           $ & $0.109\pm0.002$ & $5$ & $0.220\pm0.008$ & $0.422\pm0.008$ & $11/12$\\
                                & $(1.5,2.0)$ & $(0.15,0.20)$ & $0.945\pm0.005           $ & $0.113\pm0.002$ & $5$ & $0.197\pm0.003$ & $0.438\pm0.008$ & $5/12$\\
                                & $(1.5,2.0)$ & $(0.20,0.25)$ & $0.945\pm0.003           $ & $0.118\pm0.001$ & $5$ & $0.232\pm0.010$ & $0.457\pm0.004$ & $6/12$\\
                                & $(2.0,2.5)$ & $(0.20,0.25)$ & $0.945^{+0.015}_{-0.007} $ & $0.118\pm0.004$ & $5$ & $0.175\pm0.005$ & $0.457\pm0.015$ & $7/11$\\
                                & $(1.5,2.0)$ & $(0.25,0.30)$ & $0.945\pm0.001           $ & $0.131\pm0.001$ & $5$ & $0.335\pm0.013$ & $0.507\pm0.004$ & $12/12$\\
                                & $(2.0,2.5)$ & $(0.25,0.30)$ & $0.945\pm0.001           $ & $0.131\pm0.001$ & $5$ & $0.220\pm0.007$ & $0.507\pm0.004$ & $6/12$\\
                                & $(1.5,2.0)$ & $(0.30,0.38)$ & $0.945\pm0.002           $ & $0.135\pm0.003$ & $5$ & $0.430\pm0.020$ & $0.523\pm0.012$ & $7/(4)$\\
                                & $(2.0,2.5)$ & $(0.30,0.38)$ & $0.945\pm0.003           $ & $0.140\pm0.001$ & $5$ & $0.380\pm0.013$ & $0.542\pm0.004$ & $10/11$\\
                                & $(2.5,3.0)$ & $(0.30,0.38)$ & $0.945\pm0.002           $ & $0.142\pm0.001$ & $5$ & $0.215\pm0.010$ & $0.550\pm0.004$ & $6/11$\\
                                & $(3.0,3.5)$ & $(0.30,0.38)$ & $0.945\pm0.003           $ & $0.146\pm0.002$ & $5$ & $0.230\pm0.011$ & $0.565\pm0.007$ & $0.8/(2)$\\
                                & $(2.0,2.5)$ & $(0.38,0.48)$ & $0.945\pm0.005           $ & $0.144\pm0.001$ & $5$ & $0.630\pm0.030$ & $0.558\pm0.004$ & $14/9$\\
                                & $(2.5,3.0)$ & $(0.38,0.48)$ & $0.945\pm0.002           $ & $0.147\pm0.002$ & $5$ & $0.500\pm0.022$ & $0.569\pm0.007$ & $17/(4)$\\
                                & $(3.0,3.5)$ & $(0.38,0.48)$ & $0.945\pm0.001           $ & $0.150\pm0.002$ & $5$ & $0.300\pm0.016$ & $0.581\pm0.008$ & $29/9$\\
                                & $(3.5,4.0)$ & $(0.38,0.48)$ & $0.945\pm0.005           $ & $0.160\pm0.003$ & $5$ & $0.290\pm0.012$ & $0.620\pm0.012$ & $2/(4)$\\
                                & $(3.0,3.5)$ & $(0.48,0.58)$ & $0.945\pm0.001           $ & $0.151\pm0.002$ & $5$ & $0.480\pm0.023$ & $0.585\pm0.008$ & $0.1/(1)$\\
                                & $(3.5,4.0)$ & $(0.48,0.58)$ & $0.945\pm0.002           $ & $0.170\pm0.003$ & $5$ & $0.380\pm0.014$ & $0.658\pm0.012$ & $1/(2)$\\
                                & $(4.0,4.6)$ & $(0.48,0.58)$ & $0.945\pm0.003           $ & $0.172\pm0.005$ & $5$ & $0.320\pm0.015$ & $0.666\pm0.020$ & $0.01/(1)$\\
\hline
$\gamma^*p \rightarrow \pi^+ n$ & 1.75 & 0.25 & $0.950\pm0.005$ & $0.072\pm0.002$ & $5$ & $1.200\pm0.050$  & $0.279\pm0.008$ & $18/13$\\
                                & 1.75 & 0.31 & $0.950\pm0.010$ & $0.075\pm0.004$ & $5$ & $0.950\pm0.030$   & $0.290\pm0.015$ & $28/12$\\
                                & 2.05 & 0.25 & $0.950\pm0.010$ & $0.074\pm0.003$ & $5$ & $0.650\pm0.020$   & $0.287\pm0.011$ & $16/11$\\
                                & 2.05 & 0.31 & $0.950\pm0.003$ & $0.076\pm0.006$ & $5$ & $1.200\pm0.040$  & $0.294\pm0.023$ & $29/11$\\
                                & 2.05 & 0.37 & $0.950\pm0.010$ & $0.078\pm0.006$ & $5$ & $2.400\pm0.090$  & $0.302\pm0.023$ & $27/11$\\
                                & 2.35 & 0.31 & $0.950\pm0.004$ & $0.078\pm0.010$ & $5$ & $0.900\pm0.020$   & $0.302\pm0.039$ & $24/12$\\
                                & 2.35 & 0.37 & $0.950\pm0.010$ & $0.079\pm0.002$ & $5$ & $2.500\pm0.080$  & $0.306\pm0.008$ & $9/11$\\
                                & 2.35 & 0.43 & $0.950\pm0.006$ & $0.110\pm0.010$ & $5$ & $2.500\pm0.100$ & $0.426\pm0.039$ & $10/11$\\
                                & 2.65 & 0.31 & $0.950\pm0.010$ & $0.079\pm0.002$ & $5$ & $0.900\pm0.030$   & $0.306\pm0.008$ & $16/11$\\
                                & 2.65 & 0.37 & $0.950\pm0.008$ & $0.083\pm0.003$ & $5$ & $1.900\pm0.050$  & $0.321\pm0.012$ & $14/10$\\
                                & 2.65 & 0.43 & $0.950\pm0.010$ & $0.111\pm0.004$ & $5$ & $2.000\pm0.070$  & $0.430\pm0.016$ & $9/10$\\
                                & 2.95 & 0.37 & $0.950\pm0.010$ & $0.111\pm0.010$ & $5$ & $0.700\pm0.020$   & $0.430\pm0.039$ & $22/13$\\
                                & 2.95 & 0.43 & $0.950\pm0.004$ & $0.113\pm0.002$ & $5$ & $2.000\pm0.090$  & $0.438\pm0.008$ & $10/10$\\
                                & 2.95 & 0.49 & $0.950\pm0.012$ & $0.124\pm0.008$ & $5$ & $1.900\pm0.070$  & $0.480\pm0.031$ & $15/12$\\
                                & 3.35 & 0.43 & $0.950\pm0.010$ & $0.124\pm0.005$ & $5$ & $0.300\pm0.020$   & $0.480\pm0.019$ & $11/10$\\
                                & 3.35 & 0.49 & $0.950\pm0.010$ & $0.124\pm0.003$ & $5$ & $1.100\pm0.080$  & $0.480\pm0.012$ & $9/10$\\
                                & 3.85 & 0.43 & $0.950\pm0.006$ & $0.128\pm0.006$ & $5$ & $0.400\pm0.030$   & $0.496\pm0.023$ & $10/12$\\
                                & 3.85 & 0.49 & $0.950\pm0.003$ & $0.129\pm0.002$ & $5$ & $0.750\pm0.040$   & $0.500\pm0.008$ & $8/9$\\
                                & 3.85 & 0.55 & $0.950\pm0.003$ & $0.131\pm0.002$ & $5$ & $1.500\pm0.080$  & $0.507\pm0.008$ & $4/11$\\
                                & 4.35 & 0.55 & $0.950\pm0.002$ & $0.135\pm0.002$ & $5$ & $1.200\pm0.060$  & $0.523\pm0.008$ & $1/(4)$\\
\hline
\end{tabular}}
\end{center}
\end{table*}

\begin{figure*}[htbp]
\begin{center}
\rotatebox[origin=c]{-90}{\includegraphics[width=22cm]{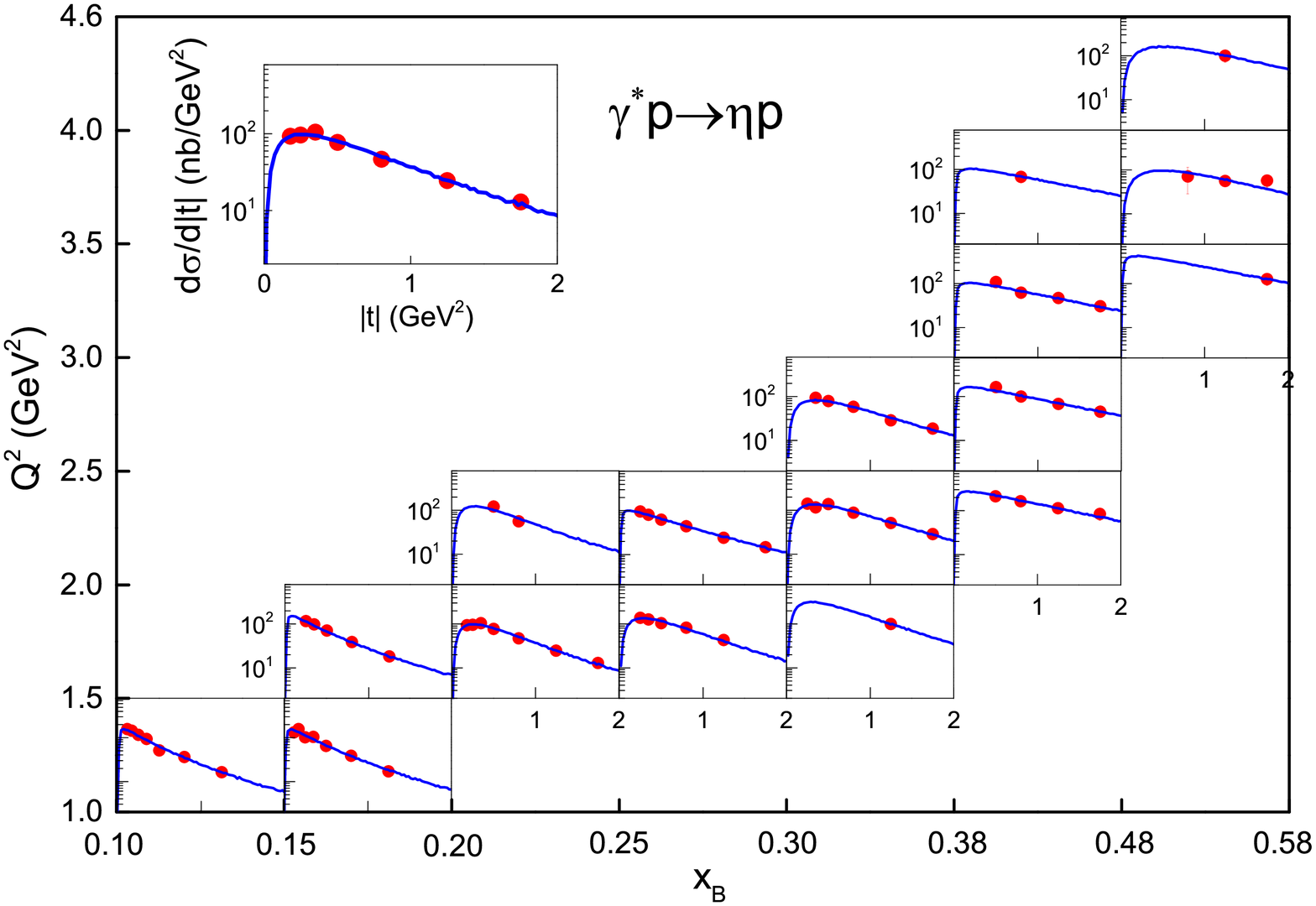}}
\end{center} \vspace{5mm}
\justifying\noindent {\small Fig. 3. The differential
cross-section $d\sigma/d|t|$ in $|t|$ of $\gamma^*p \rightarrow
\eta p$ process produced in $ep$ collisions at beam energy of 5.75
GeV in different $Q^2$ and $x_B$ ranges shown in the panels. As an
example, the sample at the top-left sub-panel shows repeatedly the
result in $1.5<Q^2<2.0$ GeV$^2$ and $0.20<x_B<0.25$. The symbols
represent the experimental data measured by the CLAS
Collaboration~\cite{16} and the curves are the statistical results
obtained as those in Figure 1.}
\end{figure*}

\begin{figure*}[htbp]
\begin{center}
\rotatebox[origin=c]{-90}{\includegraphics[width=22cm]{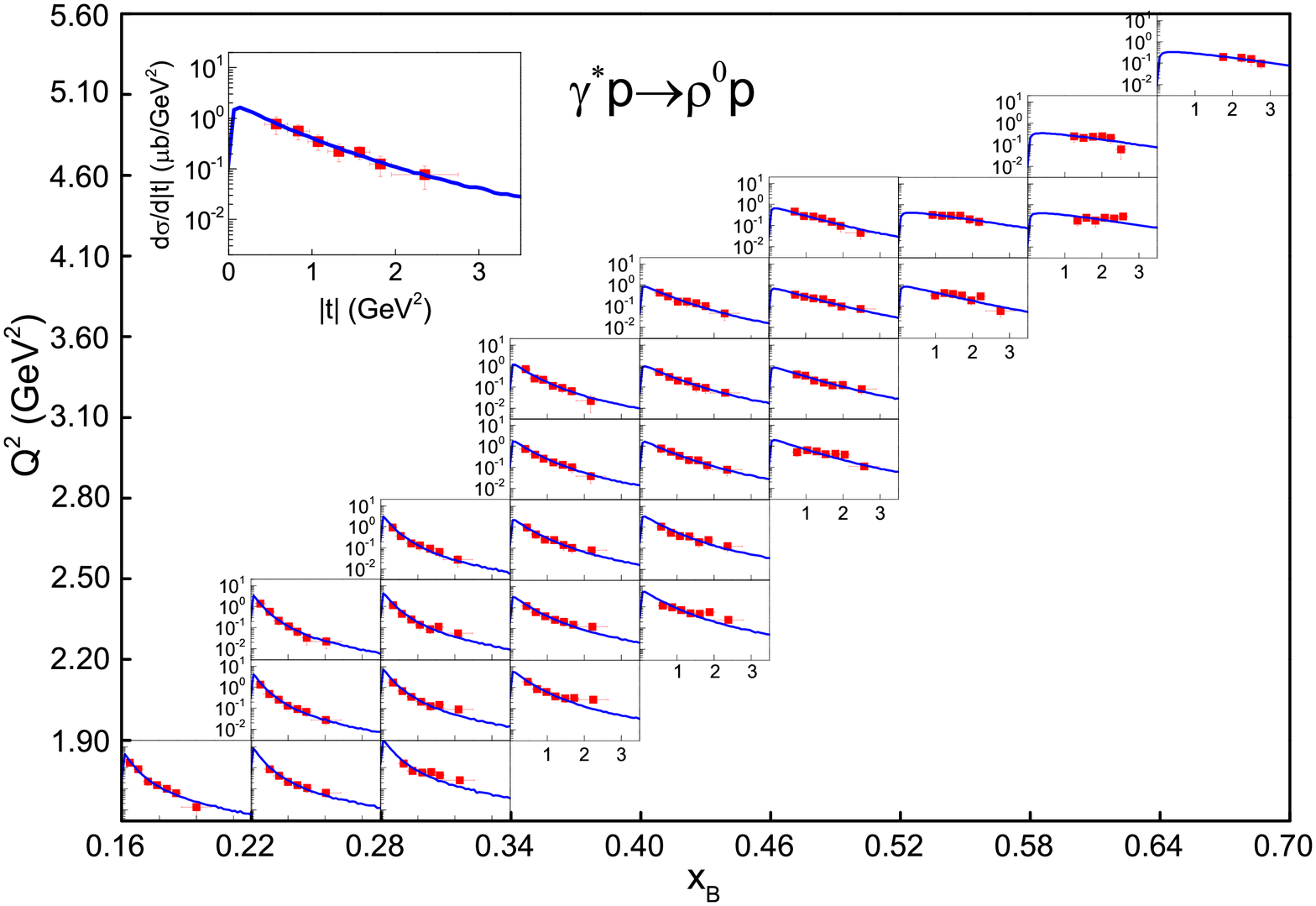}}
\end{center} \vspace{5mm}
\justifying\noindent {\small Fig. 4. The differential
cross-section $d\sigma/d|t|$ in $|t|$ of $\gamma^*p \rightarrow
\rho^0 p$ process produced in $ep$ collisions at beam energy of
5.754 GeV in different $Q^2$ and $x_B$ ranges shown in the panels.
As an example, the sample at the top-left sub-panel shows
repeatedly the result in $2.8<Q^2<3.1$ GeV$^2$ and
$0.40<x_B<0.46$. The symbols represent the experimental data
measured by the CLAS Collaboration~\cite{17} and the curves are
the statistical results obtained as those in Figure 1.}
\end{figure*}

\begin{table*}[htbp]
\vspace{.0cm} \justifying\noindent {\small Table 2. Same as Table
1, but corresponding to the curves in Figures 3 and 4.}
\vspace{-4mm}
\begin{center}
\setlength\tabcolsep{6pt}%
{\small
\begin{tabular} {cccccccccc}\\ \hline\hline  Collisions & $Q^2$ (GeV) & $x_B$ & $E_1$ (GeV) & $\langle p_t\rangle$ (GeV/$c$) & $n_s$ & $\sigma_0$ ($\mu$b) & $T_i$ (GeV) & $\chi^2$/ndof(nop)\\
\hline
$\gamma^*p \rightarrow \eta p$  & $(1.0,1.5)$ & $(0.10,0.15)$ & $0.955\pm0.005$ & $0.188\pm0.001$ & $3$ & $0.093\pm0.004$ & $0.449\pm0.003$ & $7/11$\\
                                & $(1.0,1.5)$ & $(0.15,0.20)$ & $0.955\pm0.003$ & $0.195\pm0.003$ & $3$ & $0.098\pm0.005$ & $0.466\pm0.008$ & $5/11$\\
                                & $(1.5,2.0)$ & $(0.15,0.20)$ & $0.955\pm0.002$ & $0.196\pm0.001$ & $3$ & $0.098\pm0.007$ & $0.468\pm0.003$ & $1/9$\\
                                & $(1.5,2.0)$ & $(0.20,0.25)$ & $0.955\pm0.001$ & $0.140\pm0.001$ & $5$ & $0.098\pm0.003$ & $0.557\pm0.004$ & $3/11$\\
                                & $(2.0,2.5)$ & $(0.20,0.25)$ & $0.955\pm0.010$ & $0.142\pm0.002$ & $5$ & $0.125\pm0.008$ & $0.566\pm0.008$ & $2/(2)$\\
                                & $(1.5,2.0)$ & $(0.25,0.30)$ & $0.955\pm0.002$ & $0.148\pm0.001$ & $5$ & $0.147\pm0.008$ & $0.589\pm0.004$ & $3/9$\\
                                & $(2.0,2.5)$ & $(0.25,0.30)$ & $0.955\pm0.005$ & $0.248\pm0.001$ & $3$ & $0.100\pm0.006$ & $0.592\pm0.003$ & $1/11$\\
                                & $(1.5,2.0)$ & $(0.30,0.38)$ & $0.955\pm0.010$ & $0.150\pm0.010$ & $5$ & $0.354\pm0.015$ & $0.597\pm0.039$ & $0.04/(1)$\\
                                & $(2.0,2.5)$ & $(0.30,0.38)$ & $0.955\pm0.004$ & $0.160\pm0.001$ & $5$ & $0.170\pm0.005$ & $0.637\pm0.004$ & $4/10$\\
                                & $(2.5,3.0)$ & $(0.30,0.38)$ & $0.955\pm0.003$ & $0.162\pm0.002$ & $5$ & $0.105\pm0.006$ & $0.644\pm0.008$ & $3/9$\\
                                & $(2.0,2.5)$ & $(0.38,0.48)$ & $0.955\pm0.002$ & $0.300\pm0.001$ & $3$ & $0.390\pm0.018$ & $0.716\pm0.003$ & $0.3/(1)$\\
                                & $(2.5,3.0)$ & $(0.38,0.48)$ & $0.955\pm0.002$ & $0.305\pm0.002$ & $3$ & $0.240\pm0.013$ & $0.728\pm0.005$ & $2/(4)$\\
                                & $(3.0,3.5)$ & $(0.38,0.48)$ & $0.955\pm0.002$ & $0.308\pm0.002$ & $3$ & $0.155\pm0.012$ & $0.735\pm0.005$ & $2/(4)$\\
                                & $(3.5,4.0)$ & $(0.38,0.48)$ & $0.955\pm0.002$ & $0.314\pm0.002$ & $3$ & $0.160\pm0.007$ & $0.750\pm0.005$ & $0.01/(4)$\\
                                & $(3.0,3.5)$ & $(0.48,0.58)$ & $0.955\pm0.003$ & $0.315\pm0.003$ & $3$ & $0.660\pm0.030$ & $0.752\pm0.008$ & $0.01/(1)$\\
                                & $(3.5,4.0)$ & $(0.48,0.58)$ & $0.955\pm0.005$ & $0.193\pm0.004$ & $5$ & $0.165\pm0.005$ & $0.768\pm0.016$ & $2/(3)$\\
                                & $(4.0,4.6)$ & $(0.48,0.58)$ & $0.955\pm0.004$ & $0.195\pm0.005$ & $5$ & $0.280\pm0.010$ & $0.776\pm0.020$ & $0.01/(1)$\\
\hline
$\gamma^*p \rightarrow \rho^0 p$& $(1.6,1.9)$ & $(0.16,0.22)$ & $0.960\pm0.005$ & $0.112\pm0.001$ & $3$ & $1.200\pm0.060$ & $0.274\pm0.003$ & $4/11$\\
                                & $(1.6,1.9)$ & $(0.22,0.28)$ & $0.960\pm0.004$ & $0.115\pm0.001$ & $3$ & $2.300\pm0.090$ & $0.282\pm0.002$ & $2/10$\\
                                & $(1.9,2.2)$ & $(0.22,0.28)$ & $0.960\pm0.003$ & $0.119\pm0.002$ & $3$ & $1.200\pm0.050$ & $0.291\pm0.005$ & $1/11$\\
                                & $(2.2,2.5)$ & $(0.22,0.28)$ & $0.960\pm0.003$ & $0.120\pm0.001$ & $3$ & $1.000\pm0.030$ & $0.294\pm0.003$ & $2/11$\\
                                & $(1.6,1.9)$ & $(0.28,0.34)$ & $0.960\pm0.005$ & $0.119\pm0.004$ & $3$ & $6.100\pm0.210$ & $0.291\pm0.010$ & $8/10$\\
                                & $(1.9,2.2)$ & $(0.28,0.34)$ & $0.960\pm0.007$ & $0.121\pm0.004$ & $3$ & $2.200\pm0.080$ & $0.296\pm0.010$ & $5/11$\\
                                & $(2.2,2.5)$ & $(0.28,0.34)$ & $0.960\pm0.003$ & $0.126\pm0.003$ & $3$ & $1.400\pm0.040$ & $0.309\pm0.008$ & $4/11$\\
                                & $(2.5,2.8)$ & $(0.28,0.34)$ & $0.960\pm0.010$ & $0.127\pm0.004$ & $3$ & $1.000\pm0.030$ & $0.311\pm0.010$ & $2/11$\\
                                & $(1.9,2.2)$ & $(0.34,0.40)$ & $0.960\pm0.004$ & $0.163\pm0.005$ & $3$ & $4.000\pm0.170$ & $0.399\pm0.012$ & $7/11$\\
                                & $(2.2,2.5)$ & $(0.34,0.40)$ & $0.960\pm0.020$ & $0.168\pm0.014$ & $3$ & $1.700\pm0.090$ & $0.412\pm0.034$ & $3/11$\\
                                & $(2.5,2.8)$ & $(0.34,0.40)$ & $0.960\pm0.016$ & $0.175\pm0.005$ & $3$ & $1.250\pm0.070$ & $0.429\pm0.013$ & $2/11$\\
                                & $(2.8,3.1)$ & $(0.34,0.40)$ & $0.960\pm0.006$ & $0.177\pm0.003$ & $3$ & $1.000\pm0.080$ & $0.434\pm0.008$ & $2/11$\\
                                & $(3.1,3.6)$ & $(0.34,0.40)$ & $0.960\pm0.004$ & $0.178\pm0.002$ & $3$ & $0.700\pm0.030$ & $0.436\pm0.005$ & $2/11$\\
                                & $(2.2,2.5)$ & $(0.40,0.46)$ & $0.960\pm0.020$ & $0.185\pm0.010$ & $3$ & $3.300\pm0.160$ & $0.453\pm0.024$ & $9/11$\\
                                & $(2.5,2.8)$ & $(0.40,0.46)$ & $0.960\pm0.008$ & $0.190\pm0.006$ & $3$ & $2.100\pm0.140$ & $0.465\pm0.014$ & $4/11$\\
                                & $(2.8,3.1)$ & $(0.40,0.46)$ & $0.960\pm0.020$ & $0.215\pm0.008$ & $3$ & $1.300\pm0.090$ & $0.527\pm0.020$ & $0.4/11$\\
                                & $(3.1,3.6)$ & $(0.40,0.46)$ & $0.960\pm0.010$ & $0.217\pm0.006$ & $3$ & $0.800\pm0.060$ & $0.532\pm0.015$ & $0.7/11$\\
                                & $(3.6,4.1)$ & $(0.40,0.46)$ & $0.960\pm0.015$ & $0.218\pm0.020$ & $3$ & $0.700\pm0.040$ & $0.534\pm0.049$ & $2/11$\\
                                & $(2.8,3.1)$ & $(0.46,0.52)$ & $0.960\pm0.020$ & $0.250\pm0.010$ & $3$ & $2.000\pm0.150$ & $0.612\pm0.025$ & $8/11$\\
                                & $(3.1,3.6)$ & $(0.46,0.52)$ & $0.960\pm0.002$ & $0.254\pm0.002$ & $3$ & $0.900\pm0.060$ & $0.622\pm0.005$ & $1/11$\\
                                & $(3.6,4.1)$ & $(0.46,0.52)$ & $0.960\pm0.010$ & $0.270\pm0.006$ & $3$ & $0.800\pm0.050$ & $0.661\pm0.015$ & $1/11$\\
                                & $(4.1,4.6)$ & $(0.46,0.52)$ & $0.960\pm0.009$ & $0.272\pm0.002$ & $3$ & $0.800\pm0.030$ & $0.666\pm0.005$ & $2/11$\\
                                & $(3.6,4.1)$ & $(0.52,0.58)$ & $0.960\pm0.020$ & $0.300\pm0.020$ & $3$ & $1.200\pm0.070$ & $0.735\pm0.049$ & $6/11$\\
                                & $(4.1,4.6)$ & $(0.52,0.58)$ & $0.960\pm0.020$ & $0.400\pm0.030$ & $3$ & $1.000\pm0.050$ & $0.980\pm0.073$ & $1/10$\\
                                & $(4.1,4.6)$ & $(0.58,0.64)$ & $0.960\pm0.020$ & $0.410\pm0.030$ & $3$ & $1.000\pm0.040$ & $1.004\pm0.074$ & $6/10$\\
                                & $(4.6,5.1)$ & $(0.58,0.64)$ & $0.960\pm0.010$ & $0.420\pm0.020$ & $3$ & $0.900\pm0.030$ & $1.029\pm0.049$ & $6/10$\\
                                & $(5.1,5.6)$ & $(0.64,0.70)$ & $0.960\pm0.010$ & $0.430\pm0.010$ & $3$ & $0.900\pm0.040$ & $1.053\pm0.025$ & $0.4/(4)$\\
\hline
\end{tabular}}
\end{center}
\end{table*}

In Eq. (9), $p_{3T}$ satisfies the Erlang distribution [Eq. (2)]
and we obtain it by the Monte Carlo method [Eq. (13)]. Then, the
squared momentum transfer $|t|$ is obtained statistically. In the
fitting process, two main parameters, i.e. the average transverse
momentum $\langle p_t\rangle$ contributed by each participant
parton and the number $n_s$ of participant partons are extracted
naturally. To obtain a better fit result, $E_1$ is extracted as an
insensitive parameter. In addition, a non-free parameter is the
normalization constant $\sigma_0$. The values of parameters with
selection condition ($Q^2$ and $x_B$), $\chi^2$, and the number of
degree of freedom (ndof) are listed in Table 1, where the number
of parameters is always 4 which includes $E_1$, $\langle
p_t\rangle$, $n_s$, and $\sigma_0$. In the case of ndof being less
than or equal to the number of parameters, we obtain the curve
from a ``prediction" or extrapolation based on other reasonable
fits in which the tendency of parameters is available. Meanwhile,
in these cases, the number of points (nop) is given in a bracket
to replace ndof in the table. One can see that the values of
$\chi^2$ are small in most cases, though the (necessary) dense log
scale is not easy to judge. The model results are in agreement
with the experimental data. From the values of parameters, the
average transverse momentum $\langle p_T\rangle$ and initial
temperature $T_i$ are obtained naturally.

Figure 2 presents the differential cross-section, $d\sigma/d|t|$,
in $|t|$ of $\gamma^*p \rightarrow \pi^+n$ process produced in
$ep$ collisions at beam energy of 6 GeV at different squared
photon virtuality, $Q^2= 1.75$, 2.05, 2.35, 2.65, 2.95, 3.35,
3.85, and 4.35 GeV$^2$, as well as at different Bjorken variable,
$x_B$ = 0.25, 0.31, 0.37, 0.43, 0.49, and 0.55. The symbols
represent the experimental data measured by the CLAS
Collaboration~\cite{15}. As those in Figure 1, the curves in
Figure 2 are also the statistical results of $|t|$ in which
$p_{3T}$ satisfies the Erlang distribution and is obtained by the
Monte Carlo method. The values of parameters with selection
condition ($Q^2$ and $x_B$), $\chi^2$, and nop are listed in Table
1. One can see that the model results are in agreement with the
experimental data.

Figure 3 displays the differential cross-section, $d\sigma/d|t|$,
in $|t|$ of $\gamma^*p \rightarrow \eta p$ process produced in
$ep$ collisions at beam energy of 5.75 GeV in different $Q^2$ and
$x_B$ ranges shown in the panel. As an example, the sample at the
top-left sub-panel shows repeatedly the result in $1.5<Q^2<2.0$
GeV$^2$ and $0.20<x_B<0.25$. The symbols represent the
experimental data measured by the CLAS Collaboration~\cite{16}.
The curves are the statistical results of $|t|$ in which $p_{3T}$
satisfies the Erlang distribution and is obtained by the Monte
Carlo method. The values of parameters with selection condition
($Q^2$ and $x_B$), $\chi^2$, and nop are listed in Table 2. One
can see that the model results are in agreement with the
experimental data.

Similar to Figures 1--3, Figure 4 presents the differential
cross-section, $d\sigma/d|t|$, in $|t|$ of $\gamma^*p \rightarrow
\rho^0 p$ process produced in $ep$ collisions at beam energy of
5.754 GeV in different $Q^2$ and $x_B$ ranges shown in the panel.
As an example, the sample at the top-left sub-panel shows
repeatedly the result in $2.8<Q^2<3.1$ GeV$^2$ and $0.40<x_B<0.46$
range. The symbols represent the experimental data measured by the
CLAS Collaboration~\cite{17}. The curves are the statistical
results of $|t|$ in which $p_{3T}$ satisfies the Erlang
distribution and is obtained by the Monte Carlo method. The values
of parameters with selection condition ($Q^2$ and $x_B$),
$\chi^2$, and nop are listed in Table 2. One can see that the
model results are in agreement with the experimental data.

\subsection{Parameter tendency and discussion}

In Figures 1--4, the cross-sections for $\pi^0 p$, $\pi^+ n$,
$\eta p$, and $\rho^0 p$ are fitted to show some differences in
concrete values and parameters, and some common features among
them in the tendency of curves also appear. This is caused by the
fact that different channels have different fraction ratios, and
all of them are from the same $ep$ collisions, though the
collision energies are slightly different.

The dependences of $\langle p_T\rangle$ (a, c, e, g) and $T_i$ (b,
d, f, h) on $Q^2$ in $\gamma^*p$ collisions with different emitted
channels [$\pi^0p$ (a, b), $\pi^+n$ (c, d), $\eta p$ (e, f), and
$\rho^0p$ (g, h)] are shown in Figure 5, where $\langle
p_T\rangle=n_s\langle p_t\rangle$ due to Tables 1 and 2 and the
values of $T_i$ are from Tables 1 and 2. Different symbols
represent the results for different $x_B$. One can see that
$\langle p_T\rangle$ and $T_i$ increase generally with an increase
in $Q^2$. Because $Q^2$ represents the hard scale (violent degree)
of collisions and a harder scale results in a higher excitation
degree, it is natural that larger $\langle p_T\rangle$ and $T_i$
appear at higher $Q^2$.

\begin{figure*}[htbp]
\begin{center}
\includegraphics[width=15cm]{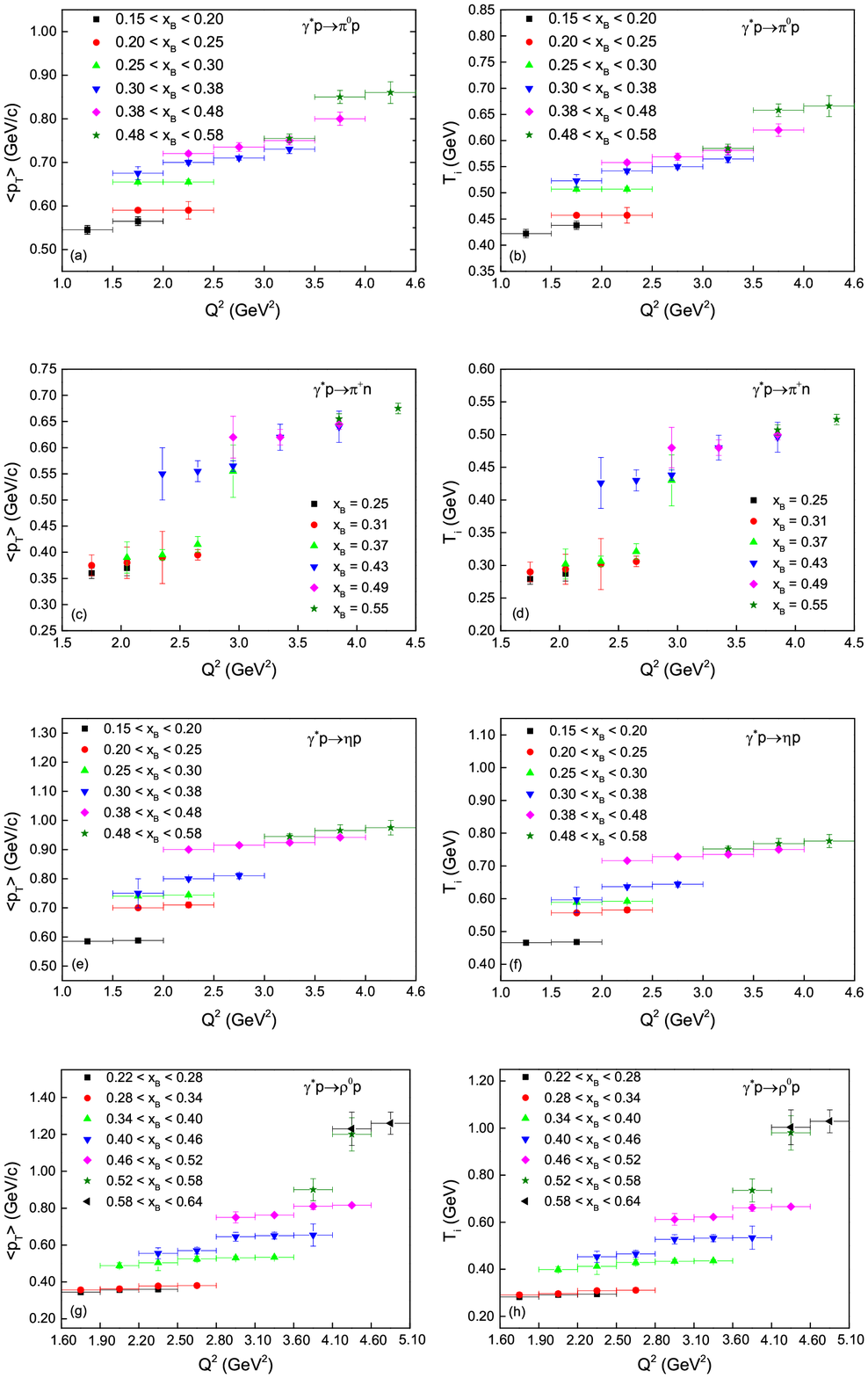}
\end{center}
\justifying\noindent {\small Fig. 5. The dependences of $\langle
p_T\rangle$ (a, c, e, g) and $T_i$ (b, d, f, h) on $Q^2$ in
$\gamma^*p$ collisions with emitted channels $\pi^0p$ (a, b),
$\pi^+n$ (c, d), $\eta p$ (e, f), and $\rho^0p$ (g, h). Different
symbols represent the results for different $x_B$.}
\end{figure*}

\begin{figure*}[htbp]
\begin{center}
\includegraphics[width=15cm]{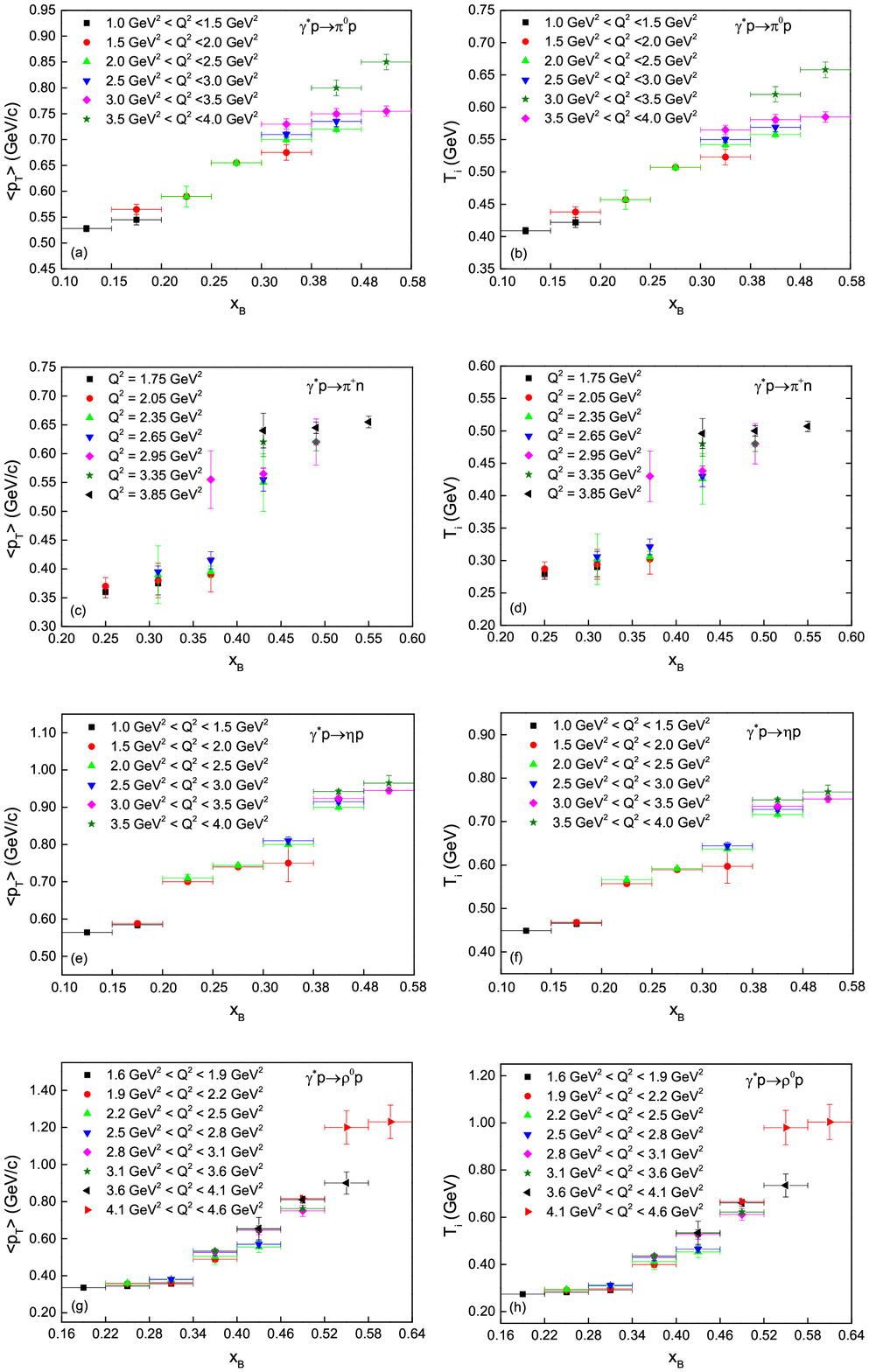}
\end{center}
\justifying\noindent {\small Fig. 6. The dependences of $\langle
p_T\rangle$ (a, c, e, g) and $T_i$ (b, d, f, h) on $x_B$ in
$\gamma^*p$ collisions with emitted channels $\pi^0p$ (a, b),
$\pi^+n$ (c, d), $\eta p$ (e, f), and $\rho^0p$ (g, h). Different
symbols represent the results for different $Q^2$.}
\end{figure*}

Figure 6 is similar to Figure 5, but it shows the dependences of
$\langle p_T\rangle$ (a, c, e, g) and $T_i$ (b, d, f, h) on $x_B$
in $\gamma^*p$ collisions with emitted channels $\pi^0p$ (a, b),
$\pi^+n$ (c, d), $\eta p$ (e, f), and $\rho^0p$ (g, h). Different
symbols represent the results for different $Q^2$. One can see
that $\langle p_T\rangle$ and $T_i$ increase generally with an
increase in $x_B$. Because $x_B\propto Q$, we may think that $x_B$
also represents the hard scale of collisions and a harder scale
results in a higher excitation degree. It is understandable that
larger $\langle p_T\rangle$ and $T_i$ appear at higher $x_B$.

In addition, $x_B$ also represents the longitudinal momentum
fraction transferred to the struck parton. In the considered
$\gamma^*p\rightarrow {\rm meson + nucleon}$ process in $ep$
collisions at given energy, the larger $x_B$ means the larger
longitudinal momentum transfer to the struck parton or the system,
and hence the more energy deposited to the system. The system
naturally stays at higher excitation degree. As a result, larger
$\langle p_T\rangle$ and $T_i$ are observed.

Generally, $\langle p_T\rangle>T_i\geq T_{ch}\geq T_0$. If the
evolution time of the system is 0, that is if the initial-state,
chemical freeze-out, and kinetic freeze-out happen simultaneously,
we have $T_i=T_{ch}=T_0$. If the evolution time is not negligible,
we have $T_i>T_{ch}>T_0$. The difference between $\langle
p_T\rangle$ and temperature is explained as the contribution of
flow effect. According to ref.~\cite{53}, in the final-state, the
expected real $T_0\approx\langle p_T\rangle/3.07$. Then, we have
the contribution of flow effect to be $\langle
p_T\rangle-T_0\approx2.07\langle p_T\rangle/3.07$. One can see
that the flow effect contributes largely to $\langle p_T\rangle$.
It is expected that the contribution of flow effect increases with
the increase of evolution time, if $\langle p_T\rangle$ is fixed
from the initial- to final-states.

From Tables 1 and 2, we note that the values of $n_s$ are 3--5 for
different channels. As the number of participant partons, $n_s$ is
constrained to be integer with uncertainty of 0. For a given
channel, $n_s$ is independent of $Q^2$ and $x_B$ in most cases.
The channel independent $n_s$ renders that the number of
participant partons is not too small or big. The number of struck
parton(s) is usually regarded as 1 or 2, which is very small. The
struck parton(s) and the partons around the struck parton(s) are
participant partons. The partons far away from the struck
parton(s) are remainder or spectator partons.

Before summary and conclusions, we would like to point out that
the discussion about the temperature and flow in this paper is
applicable. Although the multiplicity in $ep$ collisions at a few
GeV is very limited and the final particles are in a state far
from thermal equilibrium, we may use the grand canonical ensemble
for lots of events in which the number of total particles is very
large and the whole system is in a homogeneous and equilibrium
state. Therefore, the temperature used in this paper is comparable
to the freeze-out temperatures used in nucleus-nucleus collisions.
Of course, we may also regarded the temperature used here as a
fitting parameter if necessary.

The initial-temperature $T_i$ is extracted from the
root-mean-square of $p_T$, which is independent of model, though
the relation between $T_i$ and $\sqrt{\langle p_T^2\rangle}$ is
from the color string percolation method~\cite{18,19,20,21}. As
deep inelastic scattering, $ep$ collisions are head-on collisions,
and may be harder than nucleus-nucleus collisions at similar
energy per nucleon due to the fact that some non-head-on
nucleon-nucleon collisions exist in the later. As a hybrid state
of head-on and non-head-on nucleon-nucleon collisions,
nucleus-nucleus collisions may be weaker than head-on $ep$
collisions. In addition, cold spectator nuclear effect also causes
the temperature in nucleus-nucleus collisions to reduce. This
renders that $T_i$ obtained in this paper is higher than that in
nucleus-nucleus collisions.

It should be emphasized that the parameter $T_i$ reflects the
violent degree of collisions. To our knowledge, other groups and
other studies where $T_i$ is extracted for hadronic collisions is
not available at present, though $T_i$ for nucleus-nucleus
collisions is available. In terms of $T_i$, Erlang distribution,
and Monte Carlo calculation, the present work has proposed an
alternative method to describe light meson electroproduction data
obtained with the JLab-CLAS facility. Typically those data are
interpreted in terms of handbag diagram within the formalism of
generalized parton distributions, whereas here statistical
methods, that were developed for high-energy nucleus-nucleus
collisions, are applied. At least, the present work has
significance in the application of statistical methods.

\section{Summary and conclusions}

In summary, the squared momentum transfer spectra of $\pi^0$,
$\pi^+$, $\eta$, and $\rho^0$ produced in $\gamma^*p\rightarrow
{\rm meson + nucleon}$ process have been fitted by the calculated
results with the Erlang distribution which is obtained from the
multi-source thermal model and used to describe the transverse
momentum spectra of emitted particles. The squared momentum
transfer undergoes from the incident $\gamma^*$ to emitted meson,
and also equivalently from the target proton to emitted nucleon.
The model results are in agreement with the experimental data
measured by the CLAS Collaboration. The values of the related
parameters are extracted in the fitting process. The squared
photon virtuality $Q^2$ and Bjorken variable $x_B$ dependent
parameters are obtained.

With increasing of $Q^2$, the quantities $\langle p_T\rangle$ and
$T_i$ increase generally. $Q^2$ is defined as absolute value of
the squared mass of the virtual photon that is exchanged between
the electron and the target proton, and it effectively represents
the transverse size of the probe. $Q^2$ also reflects the hard
scale of collisions. A harder scale results in a higher excitation
degree of the system, and a larger $\langle p_T\rangle$ and $T_i$.
At harder scale (larger $Q^2$), the degree of equilibrium
decreases because of more disturbance to the equilibrated residual
partons in target particle, though the system is at the state of
high degree of excitation.

Similar to the tendency of $Q^2$, with an increase of $x_B$, the
quantities $\langle p_T\rangle$ and $T_i$ increase. In the
considered $\gamma^*p\rightarrow {\rm meson + nucleon}$ process,
$x_B$ represents the longitudinal momentum fraction transferred to
the struck parton. The larger $x_B$ means the larger longitudinal
momentum transfer to the system. It is natural that $\langle
p_T\rangle$ and $T_i$ are larger at larger $x_B$. In addition,
because $x_B\propto Q$, one may argue that $x_B$ also represents
the hard scale of collisions. Indeed, it is understandable that
larger $\langle p_T\rangle$ and $T_i$ appear at higher $x_B$.
\\
\\
\\
{\bf Data Availability Statement}

The data used to support the findings of this study are included
within the article and are cited at relevant places within the
text as references.
\\
\\
{\bf Author Contributions}

All authors listed have made a substantial, direct, and
intellectual contribution to the work and approved it for
publication.
\\
\\
{\bf Ethical Approval}

The authors declare that they are in compliance with ethical
standards regarding the content of this paper.
\\
\\
{\bf Disclosure}

The funding agencies have no role in the design of the study; in
the collection, analysis, or interpretation of the data; in the
writing of the manuscript; or in the decision to publish the
results.
\\
\\
{\bf Funding}

The work of Q.W. and F.H.L. was supported by the National Natural
Science Foundation of China under Grant Nos. 12047571, 11575103,
and 11947418, the Scientific and Technological Innovation Programs
of Higher Education Institutions in Shanxi (STIP) under Grant No.
201802017, the Shanxi Provincial Natural Science Foundation under
Grant No. 201901D111043, and the Fund for Shanxi ``1331 Project"
Key Subjects Construction. The work of K.K.O. was supported by the
Ministry of Innovative Development of the Republic of Uzbekistan
within the fundamental project No. F3-20200929146 on analysis of
open data on heavy-ion collisions at RHIC and LHC.
\\
\\
{\bf Conflicts of Interest}

The authors declare that there are no conflicts of interest
regarding the publication of this paper.
\\

\end{document}